\documentclass[aps,pra,twocolumn,amsmath,amssymb]{revtex4-2}

\usepackage{graphicx}
\usepackage{dcolumn}
\usepackage{bm}

\begin{document}

\title{Strongly confined atomic localization by Rydberg coherent population trapping}

\author{Teodora Kirova$^{1}$, Ning Jia$^{2\ast}$, Seyyed Hossein Asadpour$^{3}$, Jing Qian$^{4\dagger}$,Gediminas Juzeli$\overline{u}$nas$^{5}$, Hamid Reza Hamedi$^{5\ddagger}$}
\affiliation{$^{1}$Institute of Atomic Physics and Spectroscopy, University of Latvia, LV-1004, Latvia}
\affiliation{$^{2}$The Public Experimental Center, University of Shanghai for Science and Technology, Shanghai 200093, China}
\affiliation{$^{3}$Department of Physics, Iran University of Science and Technology, Tehran, Iran}
\affiliation{$^{4}$State Key Laboratory of Precision Spectroscopy, Quantum Institute for Light and Atoms,
Department of Physics, School of Physics and Electronic Science,
East China Normal University, Shanghai 200062, China}
\affiliation{$^{5}$Institute of Theoretical Physics and Astronomy, Vilnius University, LT-10257, Lithuania}


\begin{abstract}
In this letter we investigate the possibility to attain strongly confined atomic localization using interacting Rydberg atoms in a Coherent Population Trapping ($CPT$) ladder configuration, where a standing-wave (SW) is used as a coupling
field in the second leg of the ladder.
Depending on the degree of compensation of the Rydberg level energy shift induced by the van der Waals $(vdW)$ interaction, by the coupling field detuning, we distinguish between two antiblockade regimes, {\it i.e.} a partial antiblockade (PA) and a full antiblockade (FA). 
While a periodic pattern of tightly localized regions can be achieved for both regimes, the PA allows  much faster converge of spatial confinement yielding a high resolution Rydberg state-selective superlocalization regime for higher-lying Rydberg levels. 
In comparison, for lower-lying Rydberg levels the PA leads to an anomalous change of spectra linewidth, confirming the importance of using a stable uppermost state to achieve a superlocalization regime.
\end{abstract}

\email{jianing09@gmail.com}
\email{$^{\dagger}$jqian1982@gmail.com}
\email{$^{\ddagger}$hamid.hamedi@tfai.vu.lt}

\maketitle
\preprint{}

\maketitle

\begin{figure}[htbp]
\centering
\fbox{\includegraphics[width=\linewidth]{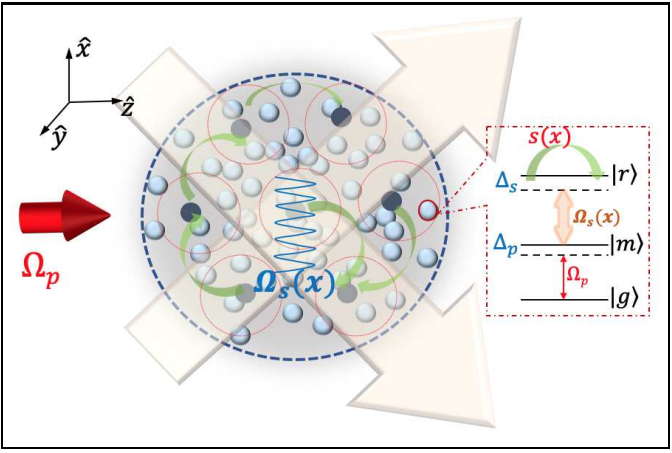}}
\caption{Schematic representation of an atomic ensemble composing of numbers of blocked Rydberg superatoms, which are coupled by a SW strong coupling field $\Omega_s(x)$ and a weak travelling wave (TW) probe field $\Omega_p$. Inset shows a three-level atom ladder configuration. $s$ describes the {\it vdWs} interacting energy between the interatomic Rydberg states.}
\label{fig:Fig1}
\end{figure}

The spacial confinement of atoms with high precision, e.g. atom localization, has been of a continuous interest in quantum mechanics, while modern tools of quantum optics have made the actual realization of such experiments possible. 
The investigations have been driven by the possibility of practical applications including nanolithography \cite{Boto:00}, laser cooling and trapping \cite{Phillips:98} and other areas of  atomic physics \cite{Adams:94}.  
\\
A high resolution localization scheme based on the phenomenon of Coherent Population Trapping $(CPT)$ \cite{Arimondo:96} was initially proposed \cite{Agarwal:06}, where extreme localization of an atom passing through the standing-wave (SW) field can be achieved, while the localization resolution can be increased via changing the relative intensity of probe and SW fields. 
A variety of methods has been developed for subwavelength localization of atoms interacting with the SW fields, for example via absorption \cite{Zhang:19, Groisman:05}, level population \cite{Papalakis:01, Ivanov:10, Mompart:09},  spontaneously emitted photons \cite{Ghafoor:11}, atom diffraction through a measurement induced grating \cite{Kunze:97} and using complex energy-level structure \cite{Hamedi:16}.
On the experimental side, using $EIT$ \cite{Harris:90} with a SW coupling laser of sinusoidally varying intensity, atom- and subwavelength atom localization were reported by the Yavuz group \cite{Yavuz:13, Yavuz:15} by utilizing the sensitivity of the atomic dark states.
\\ 
On the other hand, many problems arise when it comes to possibilities to achieve localization of Rydberg atoms, due to the difficulty to confine them in a small region with high density. 
The strong van der Waals $(vdW)$ interactions enhance the non-linear properties of Rydberg media via the phenomenon of dipole blockade \cite{Lukin:01} and open new opportunities for quantum optics and quantum information applications \cite{Saffman:10, Zheng:17}.
The latter makes the question of experimentally achievable precise localization of highly-excited Rydberg atoms an important one. 
\\
In this work we propose and analyze a theoretical scheme for the subwavelength Rydberg atom localization by applying the SW coupling field in a ladder scheme $CPT$ configuration.
We show that the Rydberg level energy shift due to the strong Rydberg-Rydberg interaction can be partially compensated by a corresponding detuning ({\it e.g.} in the regime of PA), which under certain conditions leads to spatial confinement of center-of-mass of Rydberg atoms with the precision down to subnanometer scale.
In order to predict possible experimental realizations of the localization scheme, our numerical simulations are performed under realistic parameters for atomic $^{87}Rb$.
\\
We consider an ensemble of interacting Rydberg atoms in a ladder excitation scheme shown in Fig.\ref{fig:Fig1}.
Each atom is in a three-level ladder configuration, where the ground $\left|g\right\rangle$ and middle  $\left|m\right\rangle$ states are coupled by the probe field with Rabi frequency $\Omega_p$, while the $\left|m\right\rangle$ and the Rydberg state $\left|r\right\rangle$ are connected via the coupling field $\Omega_s(x)$. 
The probe and coupling field detunings from the atomic $\left|g\right\rangle \rightarrow \left|m\right\rangle$ and $\left|m\right\rangle \rightarrow \left|r\right\rangle$ transitions are denoted by $\Delta_p$ and $\Delta_s$, respectively.
\\
We will use the Rotating Wave Approximation and assume a frozen-atom limit due to the fast operation of the experiments typically in the order of $\mu s$ \cite{Browaeys:16}. 
The Hamiltonian of the atomic system then reads $H=H_{a}+H_{af}+U_{vdW}$, where the constituting terms $H_{a}=\sum_{j}^{N}[\Delta_p \sigma^{j}_{mm}+\Delta_s \sigma^{j}_{rr}]$, $H_{af}=\sum_{j}^{N}[\Omega_p \sigma^{j}_{mg}+\Omega_s\sigma^{j}_{rm}+H.c.]$ and $U_{vdW}=\sum_{i<j}^{N}\frac{C_6}{\left| r_i-r_j\right|^6} \sigma^{i}_{rr} \sigma^{j}_{rr}$
describe, respectively, the unperturbed atomic dynamics, the atom-field coupling and the inter-nuclear $vdW$ interaction.
Here the operator $\sigma^{j}_{\alpha \beta} = \left|\alpha \right\rangle  \langle \beta|$ $(\alpha \neq \beta)$ describes atomic transitons, and $\sigma^{j}_{\alpha \alpha} = \left|\alpha \right\rangle  \langle \alpha|$ is the projection operator.
\\
Under the mean-field approximation the time evolution of the $j^{th}$ atom operator $\sigma_{\alpha \beta}^j$ is governed by the equations:
\begin{eqnarray}
\dot{\sigma}_{gg}^j&=& i \Omega_{p} \sigma_{gm}^j- i \Omega^{*}_{p}\sigma_{mg}^j +2\gamma_{gm} \sigma_{mm}^j, \nonumber\\
\dot{\sigma}_{rr}^j&=& i \Omega_{s} \sigma_{mr}^j- i \Omega^{*}_{s}\sigma_{rm}^j,  \nonumber\\
\dot{\sigma}_{gm}^j&=& (i \Delta_{p}-\gamma_{gm}) \sigma_{gm}^j + i \Omega^{*}_{p}(\sigma_{gg}^j -\sigma_{mm}^j)+i\Omega_s \sigma_{gr}^j, \label{eq:sigmadot}\\
\dot{\sigma}_{gr}^j&=& i (\Delta_{s}-s) \sigma_{gr}^j + i \Omega^{*}_{s}\sigma_{gm}^j -i \Omega^{*}_{p}\sigma_{mr}^j,\nonumber\\
\dot{\sigma}_{mr}^j&=& [i (\Delta_{s}-s-\Delta_p)-\gamma_{gm}]\sigma_{mr}^j \nonumber\\
&+&i \Omega^{*}_{s}(\sigma_{mm}^j -\sigma_{rr}^j)-i \Omega_{p}\sigma_{gr}^j,\nonumber
\end{eqnarray}
\noindent where $\Gamma_{m}=2\gamma_{gm}$, $\gamma_{mr}=\gamma_{gm}$, $\gamma_{\alpha \beta}=(\Gamma_{\alpha}+\Gamma_{\beta})/2 [\alpha$ and $\beta \in (g,m,r)]$, with $\Gamma_{m}\gg \Gamma_{r}$. Here $\Gamma_{m(r)}$ is the spontaneous decay rate of the state  $\left|m(r)\right\rangle$, while $\gamma_{gm}$ represents the dephasing rate of the $\left|g\right\rangle \rightarrow \left|m\right\rangle$ atomic transition. The parameter $s$ describes the energy shifts to the state $\left|r_{j}\right\rangle$ induced by the $vdW$ interaction with other exciting atoms usually situated beyond the blockade radius $R_b$ \cite{Tong:04}.
\\
Under the steady state condition $\dot\sigma_{\alpha\beta}(t)\equiv0$ we can derive the analytical expression for the Rydberg level population $\sigma_{rr}$ following \cite{Ma:19}:
\begin{equation}
\sigma_{rr}=\frac{I_p(I_p+I_s)}{[I_p+I_s]^2+2\Delta_{p}(\Delta_s-s)I_s+(\Delta_s-s)^2(\gamma_{gm}^2+\Delta_{p}^2+2I_p)},
\label{eq:sigmarr}
\end{equation}
where $I_{p/s}=|\Omega_{p/s}|^2$ is the laser intensity.
 When $s=0$ and $\Delta_p \ll \gamma_{gm}$,  Eq.\ref{eq:sigmarr} gives a Lorenzian lineshape with a half-linewidth of single-atom Rydberg probability $\omega=\frac{I_p+I_s}{\sqrt{\gamma_{gm}^2+\Delta_{p}^2+2I_p}}$.

An atom in Rydberg state  $\left|r\right\rangle^{i}$ would induce a $vdW$ shift of level $\left|r\right\rangle^{j}$ in another atom separated by distance $r$, which effectively translates into a two-photon detuning.
The $vdW$ interaction then blocks the excitation of all the atoms for which this shift is much larger than $w$.

The blockade radius can therefore be defined as $R_b=\left(C_6/\omega\right)^{\frac{1}{6}}$, where $C_6$ denotes the $vdW$ coefficient.
Only one atom can be excited within the blockade radius $R_b$ and the separation $r$ between the two excited atoms meets the relation $r > R_{b}$, so it is reasonable to introduce  
a short-range cut-off to the spatial integral at $R_b$ for describing the {\it vdWs} interaction given by
 $s=\int_{R_b}^{\infty}\frac{C_{6}}{\mathbf{r^{6}}} \sigma_{rr}\rho d^{3}\mathbf{r}$.
 Here $\sigma_{rr}\rho$ represents the density of excited atoms in the ensemble, $\rho$ is the atomic density and $1/\rho=4\pi R^3/3$ is the space occupied by a single atom.

With the above equations in mind and assuming $\Delta_p=0$, the magnitude of the approximated interaction $s$ becomes:
\begin{equation}
\begin{split}
s=\frac{\omega}{\xi}\sigma_{rr}=\frac{I_p(I_p+I_s)^2}{\xi [(I_p+I_s)^2+{\Delta_s}^2(\gamma_{gm}^2+2I_p)]\sqrt{\gamma_{gm}^2+2I_p}},
\label{eq:sgeneral}
\end{split}
\end{equation}
\noindent where in order to obtain $\sigma_{rr}$ from Eq. \ref{eq:sigmarr} we have used the assumption $s = 0$.
Here $\xi =\left(R/R_b\right)^3$ is treated as an adjustable parameter controlled by the atomic density and a value of $\xi
\leq 1$ means that the full blockade is attainable.
\\
Let us consider that the coupling field is a SW in the $\hat{x}$–direction $\Omega_{s}(x)=\Omega_{s0} \text{sin}\left(kx\right)$ with $\Omega_{s0}$ its peak amplitude, while the probe field is a TW.
The SW can be produced by confining the field in an optical cavity or a Fabry-Perot resonator, or by using two counter-propagating laser beams.

According to Eqs. \ref{eq:sigmarr}, when the coupling field detuning compensates the $vdW$ shift exactly by $\Delta_s=s$, the expression for the population simplifies to  $\sigma_{rr}=I_p/[I_p+I_{s}(x)]$.
A quick observation shows that at the nodes of the SW, {\it i.e.} $kx=n\pi$($n\in integers$) the Rydberg state population can robustly persist  $\sigma_{rr}\equiv 1$, resulting in a  tightly localization of ensemble of Rydberg atoms. Hence the full antiblockade (FA) case is treated as $\Delta_s\equiv s(x)$.
We also consider a PA regime in which only the Rydberg shifts at the field nodes are compensated by $\Delta_s= s(x=n\pi/k)$. In reality due to the space-dependence of interactions PA is more feasible. Equation \ref{eq:sigmarr} can be simplified in the PA regime to the form:
\begin{equation}
{\Delta_{s}}^3 \xi (2 I_p+\gamma_{gm}^2)^{3/2}+\Delta_{s} \xi I_p^2 \sqrt{2I_p+\gamma_{gm}^2}-I_p^3=0.
\label{eq:deltatilde}
\end{equation}
\noindent 
By solving the cubic Eq. \ref{eq:deltatilde} with respect to $\Delta_s$ and taking only the real root, we can obtain the value of the coupling field detuning which satisfies the  PA condition, which is:
\begin{equation}
\Delta_s=\frac{2^{1/3}W^2-2\times3^{1/3}\xi^2}
{6^{2/3}\xi W}\frac{I_p}{\sqrt{\gamma_{gm}^2+2I_p}},
\label{eq:delta_s}
\end{equation}
where $W=(9\xi^2+\sqrt{3\xi^4(4\xi^2+27)})^{1/3}.$ 
In order to explore the real performance of Rydberg localization via a SW, especially paying attention to the roles of FA and PA, we place the scheme in $^{87}\text{Rb}$ atoms where the levels $|g\rangle,|m\rangle,|r\rangle$ are represented by the actual $|5S_{1/2}\rangle,|5P_{3/2}\rangle,|60S_{1/2}\rangle$ states \cite{Pritchard:10}.
The dissipation is determined primarily by the decay rate of the excited state $|m\rangle$, given by $\gamma_{gm}/2\pi = 3.025$MHz. 
While the amplitude of the coupling field Rabi frequency is fixed at $\Omega_{s0}/2\pi=80$MHz and the wavelength $\Lambda=480nm$, the continuous probe field has a Rabi frequency $\Omega_{p}= \Omega_{s0}/\kappa$ where $\kappa=\Omega_{s0}/\Omega_{p}$ describes the relative ratio of laser amplitudes, and the wavelength 780nm.
The tunable coefficient $\kappa$ in our calculations can take different values; however in reality a large $\kappa$
value requires the probe beam to be very weak that results in a significantly long time for steady state. 
Due to the limited lifetime of the Rydberg states, we expect the best localization results to
be achieved for an optimal $\kappa$, for which the time for the system to reach steady state denoted by $T_s$ is $1/10$ of the Rydberg states lifetime.
This leaves a sufficient time for a stable localization measurement in the experiment. For example, the $|r\rangle=|60S_{1/2}\rangle$ lifetime is $226.86 \mu s$, leading to $T_s=22.68 \mu s$. 
Note that $T_s$ is increased for larger $\kappa$ values as $T_s$ depends on the absolute values of the Rabi frequencies. 
Since $T_s$ is restrained by the Rydberg level lifetime, it is necessary to find out a best $\kappa$ value for spatially localizing the atoms. 

\begin{figure}[htbp]
\centering
\fbox{\includegraphics[width=\linewidth]{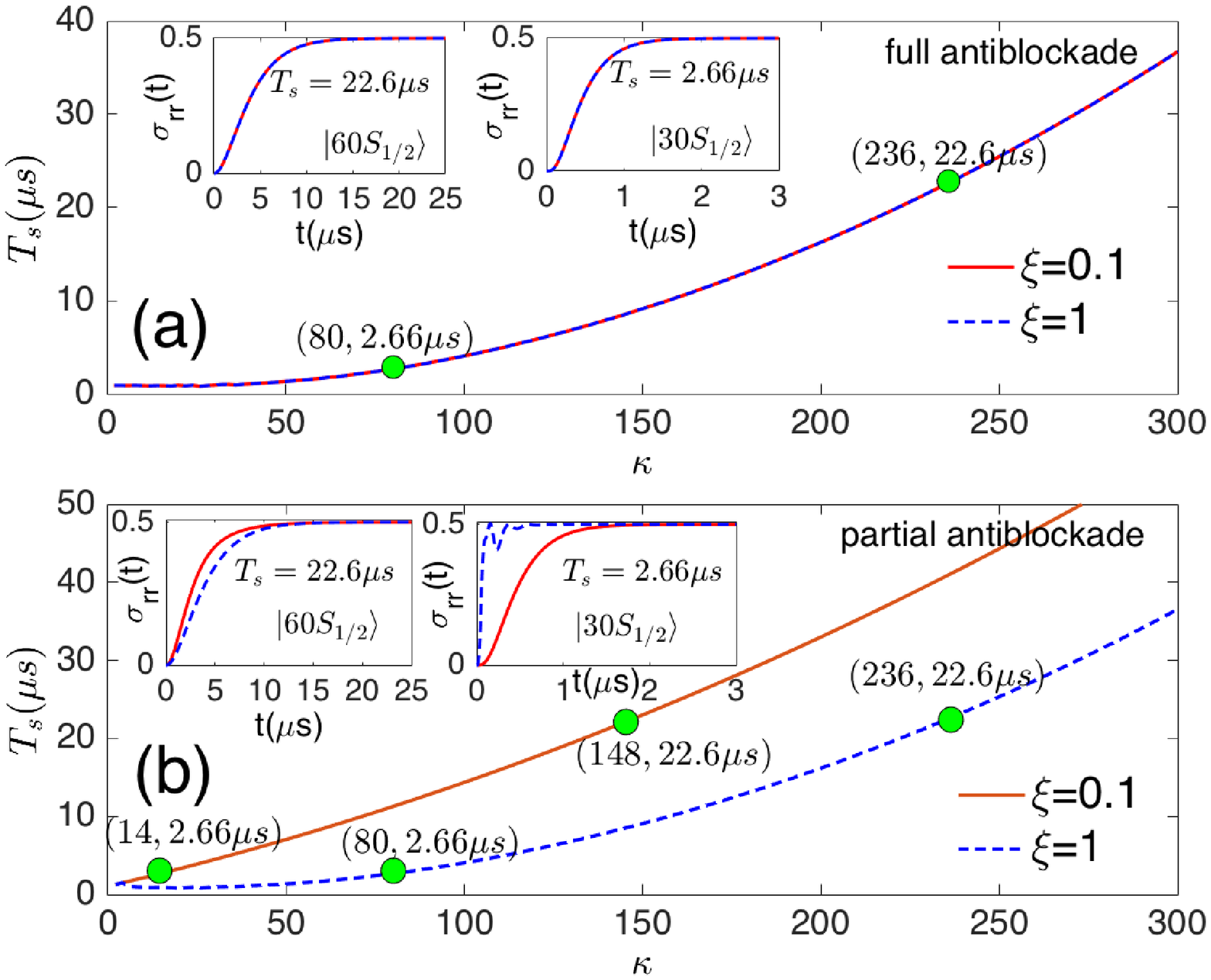}}
\caption{The relation between the steady-state time $T_s$ and $\kappa$ for the Rydberg energy levels $|r\rangle =|60S_{1/2}\rangle$ and $|30S_{1/2}\rangle$ in the cases of (a) FA: $\Delta_s\equiv s$ and (b) PA: $\Delta_s = s(x=n\pi/k)$. Different $\xi$ values are comparably displayed for $\xi=0.1$(red solid) and $\xi=1.0$(blue dotted). Insets show the transient response of population $\sigma_{rr}(t)$ for two real Rydberg levels, accordingly. }
\label{fig:Fig2}
\end{figure} 

In order to estimate the optimal $\kappa$ values, we plot in Fig. \ref{fig:Fig2} the $\kappa-T_s$ relation by directly solving the equations of motion (\ref{eq:sigmadot}). 
For comparison we also introduce a lower-lying Rydberg level $|r\rangle=|30s_{1/2}
\rangle$ with its lifetime about 26.6$\mu$s, giving $T_s=2.66\mu$s for a steady localization. 
A transition of $|5S_{1/2}\rangle\to|5P_{1/2}\rangle\to|30S_{1/2}\rangle$ is also possible in practice \cite{Han:16}.
From Fig.\ref{fig:Fig2}(a) it is clear that in the case of FA, a steady localization can be ensured by $\kappa=236$ and $\kappa=80$ for Rydberg energy levels $|r\rangle =|60S_{1/2}\rangle$ and $|30S_{1/2}\rangle$, respectively, as highlighted by green dots. 
Changing the atomic density $\xi$ does not affect the steady time $T_s$. 
The detailed transient responses are shown in the insets.
Shown in Fig.\ref{fig:Fig2}(b) for the PA regime, a smaller $\xi$ (equivalent to a stronger atomic density), increases the time $T_s$ due to the imperfect energy-shift compensation. 
Hence, the optimal $\kappa$ has to be significantly reduced to reach the same $T_s$ , which may cause a broadening of localization peaks as confirmed by Fig.\ref{fig:Fig4}(c).

\begin{figure}[htbp]
\centering
\fbox{\includegraphics[width=\linewidth]{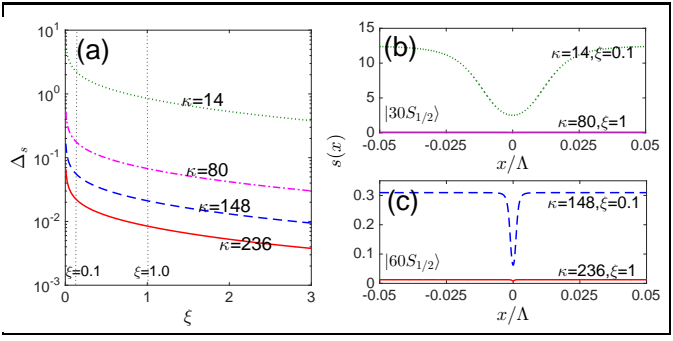}}
\caption{(a) The relation of $\Delta_s$ and $\xi$ for values of $\kappa$ for the Rydberg energy levels $|r\rangle =|60S_{1/2}\rangle$ or $|30S_{1/2}\rangle$ determined by Fig \ref{fig:Fig2}. (b-c) show the relation of the Rydberg shift $s$ and position $x$ for $\xi=0.1$ and $\xi=1.0$ for the corresponding $\kappa$; (b) Rydberg energy level $|r\rangle =|30S_{1/2}\rangle$; (c) Rydberg energy level $|r\rangle =|60S_{1/2}\rangle$.$\gamma_{gm}$ is treated as the frequency unit.}
\label{fig:Fig3}
\end{figure}

\begin{figure}[htbp]
\centering
\fbox{\includegraphics[width=\linewidth]{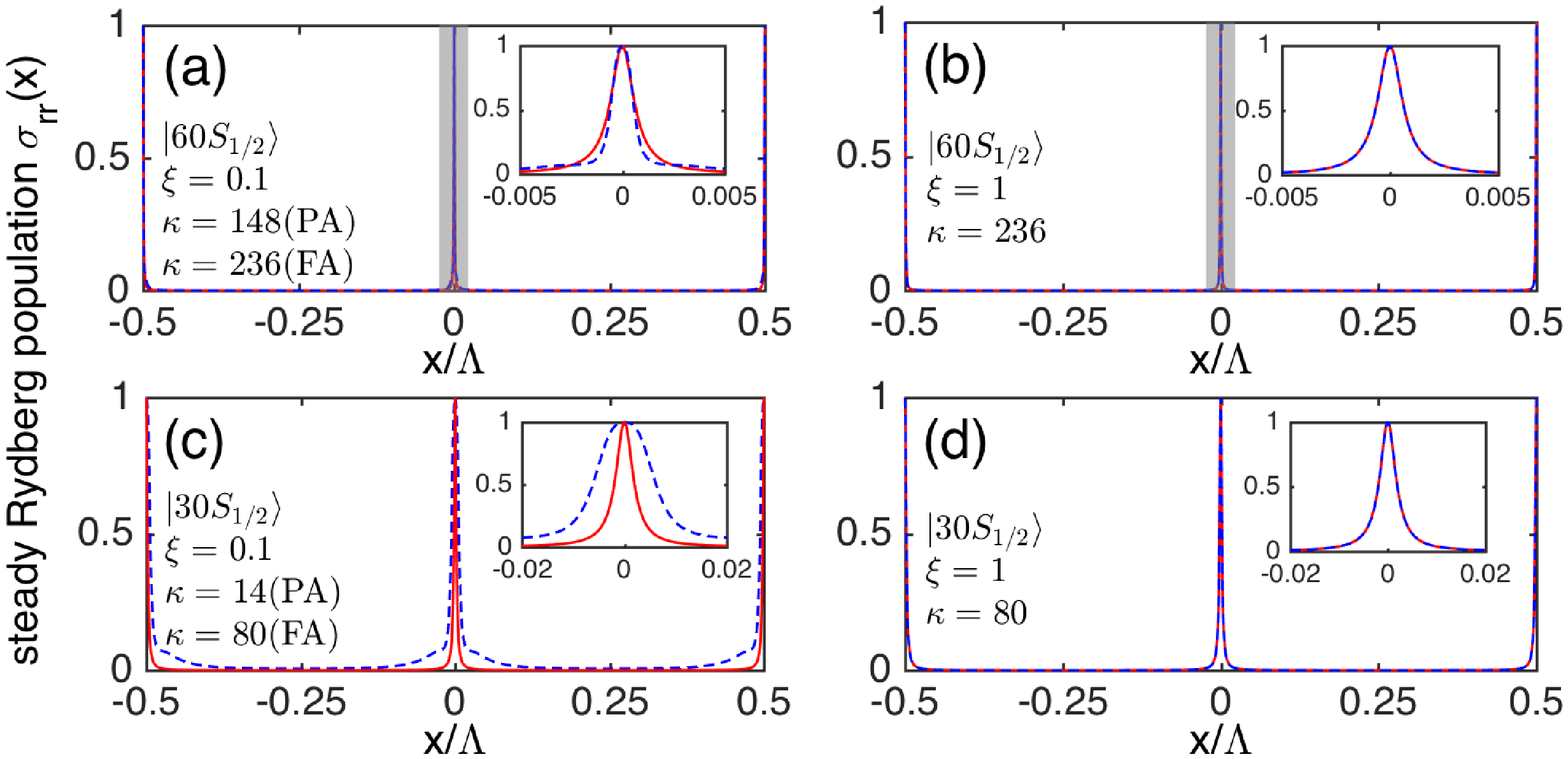}}
\caption{Steady state Rydberg population as a function of $x/\Lambda$ with parameters: $\Omega_{s0}/2\pi=80MHz$, $\Omega_p=\Omega_{s0}/\kappa$, $\gamma_{gm}/2\pi=3.025MHz$, $\Delta_p=0$.  
Red lines-FA ($\Delta_s\equiv s(x)$), blue dashed lines-PA under the condition $\Delta_s= s(x=n\pi/k)$. Parameters $\xi$ and $\kappa$ are described in the figure. Insets show a zoom-in around $x/\Lambda=0$, clearly showing the difference between the two cases.}
\label{fig:Fig4}
\end{figure}

Next, in order to elucidate the role of the atomic density $\xi$, we study quantitatively the distribution of the blocked energy $s$ in space. 
Based on Eq.(\ref{eq:delta_s}), Fig. \ref{fig:Fig3}(a) illustrates the dependence of $\Delta_s$, on $\xi$ and for different optimal $\kappa$ values as obtained by Fig.\ref{fig:Fig2}.
Intuitively, $\Delta_s$ decreases with $\xi$ because a bigger atomic density with $R>R_b$ would cause the interatomic interaction weaker.
Moreover, a larger $\kappa$ is accompanied by a smaller $\Omega_p$, lowering the Rydberg-state probability the same as the Rydberg shift $s$. Figures \ref{fig:Fig3}(b-c) show the spatial dependence of the Rydberg shifted energy $s(x)$. 
Except that $s(x)$ is significantly decreased if $\kappa$ is enhanced due to a weaker probe intensity, it is clear that a smaller $\xi$(=0.1) causes a large dip at the localized points $x=n\pi/k$ of the spatial profile of $s$.
For the case with $|60S_{1/2}\rangle$ and in the regime of PA, as shown in (c)[blue-dashed curve], the shift $s$ is overcome by a suitable two-photon detuning $\Delta_s$ at the nodes $x=n\pi/k$, resulting in an effective two-photon resonance at those positions. 
Yet, out of the nodes $x\neq n\pi/k$,  $s$ sustains a higher level which can not be overcome by $\Delta_s$, causing the steady solution $\sigma_{rr}$ to decrease significantly.
The role of PA is not dominant if $\xi$ is larger ({\it e.g.} $\xi=1$, red-solid curve), as in this case the energy $s$ remains almost the same for all positions $x=n\pi/k$ and $x\neq n\pi/k$.
On the other hand, the FA regime can self-consistently compensate the shifted energy $s(x)$ at all spatial positions $x$, preventing to improve the spectra linewidth by different atomic densities. 

Figure \ref{fig:Fig4} presents the simulations of the state-selective localization protocol by measuring the population distribution of the Rydberg state. Clearly seen in this figure is the effect of the two parameters $\kappa$ and $\xi$, as well as the difference between the FA and PA regimes.
Adopting the optimal $\kappa$ value $\kappa=236$ for $|60S_{1/2}\rangle$, as obtained by Fig.\ref{fig:Fig2}, the Rydberg population $\sigma_{rr}$ reaches its maximal value of $1.0$ at the notes of SW, creating a periodic pattern of tightly localized regions for both PA and FA regimes.   
While $\xi=1$ leads to matching
of the localization lines in both regimes, as shown in Fig. \ref{fig:Fig4} (b) (because the energy $s$ almost persists a constant for all positions),
the PA regime improves the spatial confinement significantly by reducing the spectral line-width with respect to the FA when $\xi=0.1$ (see Fig. \ref{fig:Fig4} (a)).

The reason could be understood as follows. For the case of FA by using the condition $\Delta_s\equiv s(x)$ in Eq. \ref{eq:sigmarr} we obtain:
$\sigma_{rr}^{FA}=1/(1+\kappa^2 sin^2(kx))$, where we have included the spatial dependence of the coupling field as $\Omega_{s}(x)=\Omega_{s0} sin(kx)$.
At the same time,  the full width at half maxima ($FWHM$) of the localization spectral lines is given by $a^{FA}=\arcsin{(1/\kappa)} \Lambda/ \pi$, leading to $0.647nm$ for the parameters in Fig. \ref{fig:Fig4}(a).
An analytical solution in the case of PA is impossible to obtain since the dependence of the parameter $s$ on $x$ is rather complicated. 
Therefore, for the parameters used in Fig. \ref{fig:Fig4}(a) we can make only numerical estimates, giving a $FWHM$ of $0.56nm$.

If considering a lower-lying Rydberg state $|r\rangle=|30s_{1/2}\rangle$, the localization is accompanied by a broadening of the spectral line-widths, e.g. the localization peak is not as sharp with respect to the case with the higher-lying Rydberg state $|r\rangle=|60s_{1/2}\rangle$, as depicted in Figs. \ref{fig:Fig4}(c) and (d). Moreover, an anomalous behaviour of spectra linewidth is observed now between the FA and PA regimes, {\it i.e.} the localization peaks are wider now for the PA regime, with a FWHM of $\sigma_{rr}$ to be as wide as $5.93nm$. This could be due to the wider variation of interaction energy $s$ around the localization points [see Fig. \ref{fig:Fig3}(b), dotted-green line]. Such a comparison confirms the importance of using a stable uppermost state to achieve a superlocalization regime.

In conclusion, we have studied theoretically the possibility to achieve strong localization of interacting Rydberg atoms arranged in a 3-level ladder Coherent population Trapping configuration. 
The ground and middle states are coupled by a travelling wave probe field, while the middle and Rydberg states are connected via a standing wave coupling field.
We have distinguished between two antiblockade regimes, a full antiblockade and a partial antiblockade, in which the detuning of the coupling field compensates the parameter $s$ fully or partially, only at the nodes of the standing wave coupling field.
The parameter $s$ describes the energy shift to the Rydberg state induced by the $vdW$ interaction with other excited atoms.
It has been numerically shown that the sharpest Rydberg state-selective localization is achieved under the partial blockade condition, when utilizing higher-lying Rydberg levels, for example $|r\rangle =|60S_{1/2}\rangle$, giving population equal to unity at the nodes of the standing wave, along with a $FWHM$ of $0.56nm$.
The spatial confinement is possible also when considering a lower-lying Rydberg state, e.g. $|r\rangle =|30S_{1/2}\rangle$. 
Yet, the use of such a shorter lived Rydberg state is accompanied by a spectral line widening and a loss of the localization sharpness. 
Thus, the use of more stable higher Rydberg states is preferable for the superlocalization.
Our findings are relevant in the emerging field of subnanometer localization of atoms \cite{Copeland:18}, enabling novel applications in optical microscopy techniques in real experiments.
The theoretical scheme for achieving subnanometer localization of Rydberg atoms presented in this Letter brings us one-step closer to the target of advancing the practice of localization microscopy.
Future investigations of the sub-wavelength atom localization are planned to involve more complicated level configurations, including the inverted $Y$ and the diamond schemes, and will address the situations when multiple laser fields carrying the orbital angluar momentum are applied. 

\medskip
\noindent\textbf{Funding.} This work was supported by the NSFC (No. 11474094, No. 11104076); and The Science and Technology Commission of Shanghai Municipality (No. 18ZR1412800) for J. Q. 

\medskip

\noindent\textbf{Acknowledgements.} This work was supported by STSM Grants for H.R.H. and T.K. from COST Action CA16221.

\medskip

\noindent\textbf{Disclosures.} The authors declare no conflicts of interest.

\bibliography{sample}

\end{document}